## TITLE PAGE

# Lord's 'paradox' explained: the 50-year warning on the use of 'change scores' in observational data

Peter W. G. Tennant[1,2], Georgia D. Tomova[3], Eleanor J. Murray[4], Kellyn F. Arnold[1], Matthew P. Fox[4], Mark S. Gilthorpe[5]

[1]*Leeds Institute for Data Analytics, University of Leeds, Leeds, UK*
[2]*School of Medicine, University of Leeds, Leeds, UK*
[3]*Centre for Longitudinal Studies, University College London, London, UK*
[4]*Departments of Epidemiology and Global Health, Boston University, Boston, MA, USA*
[5]*Obesity Institute, Leeds Beckett University, Leeds, UK*

## Corresponding author

*Name:* Peter W. G. Tennant, PhD

*Address*: Leeds Institute for Data Analytics,
Level 11 Worsley Building, Clarendon Way,
Leeds, LS2 9NL, UK.

*Email*: **P.W.G.Tennant@leeds.ac.uk**

*LinkedIn*: **peterwgtennant**

*Bluesky*: **pwgtennant.bsky.social**

## Funding

This work received no specific funding.

## Conflicts of interest

PWGT is a director of Causal Thinking Ltd and MSG is a director of Causal Insights Solutions Ltd, both of which provide causal inference research and training. Both companies and their directors may benefit from any study that demonstrates the value of causal inference methods.

## ABSTRACT

In 1967, Frederick Lord posed a conundrum that has confused scientists for over half a century. Subsequently named Lord's 'paradox', the puzzle centres on the observation that two different approaches to estimating the effect of an exposure on the 'change' in an outcome can produce radically different results. Approach 1 involves comparing the mean 'change score' between exposure groups and Approach 2 involves comparing the follow-up outcome between exposure groups conditional on the baseline outcome.

Resolving this puzzle starts with recognising the three reasons that a variable may change value: (A) 'endogenous change', which represents autocorrelation from baseline, (B) 'random change', which represents change from transient random processes, and (C) 'exogenous change', which represents all non-endogenous, non-random change and contains all change that is potentially modifiable by other baseline variables.

In observational data, neither Approach 1 nor Approach 2 can reliably estimate the causal effect of an exposure on 'exogenous change' in an outcome. Approach 1 is susceptible to diluted or opposite-sign estimates whenever the exposure causes, or is caused by, the baseline outcome. Approach 2 is susceptible to inflated estimates due to measurement error in the baseline outcome and time-varying confounding bias when the baseline outcome is a mediator. The measurement error can be reduced with multiple measures of the baseline outcome, and the time-varying confounding can be reduced using g-methods.

Lord's 'paradox' offers several enduring lessons for observational data science including the importance of a well-defined research question and the problems with analysing change scores in observational data.

## INTRODUCTION

In 1967, Frederick Lord posed a conundrum that has confused scientists for over half a century.[1] The puzzle - subsequently named Lord's 'paradox' - concerns the common setting of analyses of change in an outcome measured at two time points. At the heart of the puzzle is Lord's observation that two common approaches to examining the effect of an exposure on the 'change' in the outcome can return radically different results when the exposure is not assigned at random. Many articles have since attempted to resolve the 'paradox', offering a range of explanations.[2–7] To the casual reader, these discussions may appear as little more than statistical curiosities, but Lord's 'paradox' has a surprising and enduring relevance to contemporary data science.

In the following, we introduce and explore the 'paradox' and discuss the lessons of this 50-year puzzle. We begin by introducing the 'paradox', as it was first described, and examine the potential research questions being posed using a contemporary causal inference perspective. We then delve into the heart of the puzzle by examining the challenges with analyses of change in observational data by considering the definition of 'change', the outcome of interest, and the target estimand. We resolve the puzzle by explaining how the two approaches are inherently prone to biases that act in opposite directions. We conclude by recreating Lord's original example in simulated data to illustrate the key lessons.

## LORD'S 'PARADOX' INTRODUCED

Lord first described his eponymous 'paradox' in a short note published in *Psychological Bulletin* in 1967.[1] The piece described a fictional examination of the effects of sex and diet on the change in weight among university students between the start and end of term, in which two fictional statisticians using different approaches offer radically different conclusions. Lord's original example has several quirks that can potentially distract from a puzzle that generalizes far beyond the setting of sex differences in weight change. The 'paradox' has hence often been repackaged to offer clearer and/or more dramatic examples.[5,6,8,9] We nevertheless focus on Lord's original example because some of the peculiarities are themselves educational.

Lord established the puzzle as follows:

> *'A large university is interested in investigating the effects… of the diet provided in the university dining halls and any sex difference in these effects'.*
>
> *'The weight of each student at the time of his [or her] arrival in September and… the following June are recorded'.*
>
> *'At the end of the school year, the data are independently examined by two statisticians'.*

Lord then described how each statistician approached the question and came to different conclusions:

> *'The first statistician examines the mean weight of the girls at the beginning of the year and at the end of the year and finds these to be identical… He finds the same to be true for the boys. He concludes… there is no evidence of any interesting effect of the school diet… on student weight… [and] no evidence of any differential effect on the two sexes…'*
>
> *'The second statistician… decides to do an analysis of covariance [ANCOVA]… He determines that the slope of the regression line of final weight on initial weight is essentially the same for the two sexes… He finds that the difference between the intercepts is statistically highly significant. The*

> *second statistician concludes… the boys showed significantly more gain in weight than the girls when proper allowance is made for differences in initial weight.'*

Lord also provided a visual summary of the puzzle, adapted in **Figure 1**.

**Figure 1**

Reproduction of the diagram used by Fredrick M Lord to explain his eponymous 'paradox' in 1967.

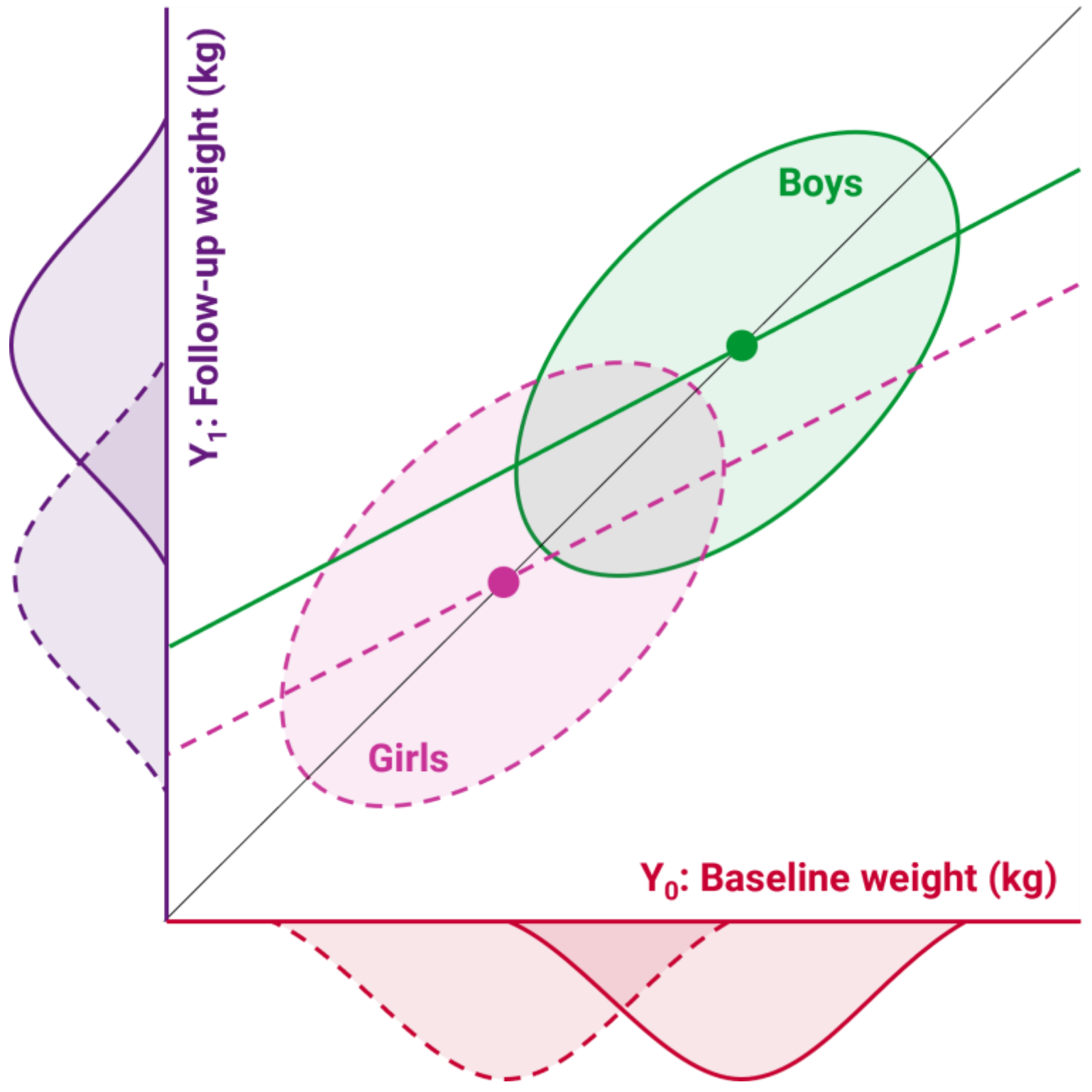

*Lord annotated the diagram as follows [with our amendments in square brackets]: 'The two ellipses represent separate scatterplots for the boys [depicted in green with solid lines] and the girls [depicted in pink with dashed lines]… The distributions of initial weight [$Y_0$, shown in red] are indicated at the… [bottom]… of the diagram [and the] distributions of final weight [$Y_1$, shown in purple] are indicated on the left... People falling on the solid 45° line… are people whose initial and final weight are identical… The two regression lines [from the second statistician's analysis] are shown [in solid green for boys and dashed pink for girls].'*

We can summarize the two approaches taken by the two statisticians as follows:

## Approach 1: 'Change scores' analyses

'**Change score**' (also known as 'gain score' or 'difference score') values ($\Delta Y = Y_1 - Y_0$) are calculated by subtracting the 'baseline' outcome measure ($Y_0$) from the 'follow-up' outcome measure ($Y_1$) and compared between exposure groups (denoted by $X$). This may be done by comparing the mean change score between the groups, or by analysing the individual change score values using one-way analysis of variance (ANOVA), a regression model of the following form, or similar:

$$E[\Delta Y|X] = E[Y_1 - Y_0|X] = \alpha_0 + \alpha_1 X \qquad \text{Eq. 1}$$

## Approach 2: 'Follow-up conditional on baseline' analyses

Values of the follow-up outcome are compared between exposure groups, conditioning on the baseline outcome. This may be done using **one-way analysis of covariance** (ANCOVA), a regression model of the following form, or similar:

$$E[Y_1|X,Y_0] = \beta_0 + \beta_1 X + \beta_2 Y_0 + (\beta_3 X \cdot Y_0) + \cdots \quad \text{Eq. 2}$$

Note Eq. 2 is equivalent to a simple ANCOVA when the interaction term ($\beta_3 X \cdot Y_0$) is omitted.

In Lord's original example, the two approaches return paradoxically different results; with the analysis of change scores suggesting that sex and the dining hall has *no* effect on 'change' in weight and the analysis of follow-up conditional on baseline suggesting that sex and the dining hall *do* contribute to 'change' in weight (indicated in Figure 1 by the differing intercepts). Subsequent examples - in both imaginary and real data - have shown the two approaches can even generate sign-discordant results.[9]

Unfortunately, the literature does not provide a clear explanation of which, if either, approach is most generally appropriate. In later writing, Lord favours analyses of change scores, noting that, '*it is quite clear that analysis of covariance is not going to provide us with a good answer*'.[8] Locascio and Cordray (1983) agree, '*[Change score analyses]… provided a correct assessment of treatment effects*'.[2] Others argue the opposite. London and Wright (2001), for example, conclude that, '*Usually the ANCOVA approach will be preferred*'.[10]

Most authors, however, do not specifically favour either approach, instead concluding that they simply answer different questions. Pearl (2016), for example, states that, '*both statisticians were… correct… [but] each estimated a different effect*'.[5] Holland and Rubin (1986) more cautiously conclude that, '*if both statisticians made only descriptive statements, they would both be correct… [but for] causal statements, neither would be correct or incorrect because [of] untestable assumptions*'.[3]

But is this satisfactory? Lord posed a simple question involving just three variables (a baseline exposure, and two measures of an outcome) yet showed that radically different results can be obtained using two ostensibly reasonable approaches. Although fault and ambiguity can be found in various aspects of Lord's original description, it seems insufficient to conclude that the two approaches are simply answering different questions. As Hand (1994) puts it, '*this deconstruction does not completely resolve the issue [because] it is still necessary for the researcher to be clear about which problem should be solved*'.[11]

## EXAMINING THE RESEARCH QUESTION

A critical problem with Lord's original example is the lack of clarity around the exact aim of the research. In his original note, Lord describes the research as being '*interested in investigating the effects… of the diet provided in the university dining halls and any sex difference in these effects*'. From this, it seems unambiguous that the exposure of interest is the *diet*, perhaps determined by *dining hall*. Unfortunately, the clarity ends there. For one thing, '*the diet provided in the university dining halls*' is not a completely defined exposure as no comparator is provided; did the research seek to compare the effect of different diets provided in different halls? Or did they seek to compare the diet provided by all halls with the diet provided in other settings?

Regardless, after the introduction, Lord's original note does not consider either the diet or the dining halls any further.[1] Both fictional statisticians and the rest of the note instead focussed on 'sex', suggesting 'sex' was in fact the primary exposure of interest. Sex is, however, a problematic exposure, reflecting an ill-defined combination of biological, behavioural, sociological, and cultural factors,[12] few of which seem relevant to '*the diet provided in the university dining halls*'.

Some of this confusion may be attributable to cultural changes in university dormitory and catering practices since the 1960s. At the time of Lord's original note, most university students would likely have been living in single-sex dormitories and eating in single-sex dining halls. Since these halls may have provided different diets, it seems plausible that the true interest was in comparing the effect of the different diets provided in the male and female dining halls. Alas, with only two halls serving exactly two diets, the assignment of which is completely determined by sex, it is impossible to separate the effects of sex, hall, and/or diet.[3,13] We must therefore conclude that sex is the only possible exposure, while noting that it will include the downstream effects of hall and diet.

These uncertainties highlight the importance of starting any causal analysis with a well-defined causal question.[14] However, even when the question is framed more clearly it does not resolve the puzzle. For example, Pearl (2016) provides both statisticians with ostensibly clear estimands stating that, '*The first statistician estimated the total effect of gender on gain [in weight]… while the second statistician estimated the direct effect*',[5] but, as we outline below, this description overlooks the ambiguity of this first 'gain' or 'change' estimand.

## THE GENERAL PUZZLE WITH ANALYSES OF CHANGE

At the heart of Lord's 'paradox' lies the larger puzzle concerning the use of 'change scores' in observational data. In experimental data, where the exposure has been randomized, both change score analyses and follow-up conditional on baseline analyses converge on the same estimand, although the latter approach is generally favoured for its greater precision.[15] Unfortunately, in observational data, there is no such consensus, either in the values obtained or the opinions on which approach is more appropriate. For example, Van Breukelen (2006) argues that '*in nonrandomized studies of pre-existing groups… ANOVA of change seems less biased than ANCOVA*',[16] while Glymour et al (2005) conclude, '*in some cases, change-score analyses without baseline adjustment provide unbiased causal effect estimates when baseline-adjusted estimates are biased*'.[17] Conversely, Senn (2006) argues '*although many situations can be envisaged where ANCOVA is biased it is very difficult to imagine circumstances under which … [a change score analysis] would then be unbiased*'.[18] Similarly, Shahar and Shahar (2010) argue that '*modelling the change between two time points is justified only in a few situations*'.[19] It is difficult to reconcile these radically different positions by simply claiming that the two approaches answer different questions.

## DEFINING 'CHANGE'

The idea of estimating the effect of an exposure on the change in an outcome seems like a simple aim, but it is complicated by a lack of formal consideration or consensus on the meaning of 'change'. In the introduction to Shahar and Shahar (2010), the authors allude to this by quoting two peer reviewers offering equally certain but opposing views on whether 'change' is a cause or consequence of follow-up.[19]

In our previous piece on analyses of change scores in observational data, we outline three reasons why a variable changes value over time.[20] The first, '**endogenous change**' (or 'determined change'), represents the realisation of all pre-determined processes that act through the baseline instance of a time-varying variable. In a causal diagram, 'endogenous change' is represented by the arc between the baseline and follow-up instance of a repeated variable and essentially captures its autocorrelation (**Figure 2**).

The second, '**random change**', represents the contribution of all transient random sources of variation, including transient classical measurement error. This can be depicted in a causal diagram by adding random variation, or 'error', terms (e.g., $R_{Y0}$, $R_{Y1}$), which combine with the underlying latent (or 'true') versions of each variable (e.g., $L_{Y0}$, $L_{Y1}$) to form the instantaneous, or 'measured', versions (e.g., $Y_0$, $Y_1$) (**Figure 2**).[21] The overall random change (e.g., $R_Y$) is then determined from the difference between the random variation in the measured versions of baseline and follow-up (i.e., $R_Y = R_{Y1} - R_{Y0}$) (**Figure 2**).

Finally, '**exogenous change**', represents all non-transient and non-random reasons for a variable changing value over time beyond what has been functionally pre-determined by and through the baseline instance. It can be depicted in a causal diagram as a latent process (e.g., $C_Y$) that causes the follow-up measure and is unrelated to the baseline measure (**Figure 2**).

**Figure 2**

Directed acyclic graph showing the three reasons why a variable ($Y$) can change value over time, also depicting the (derived) change score variable ($Y_1 - Y_0$).

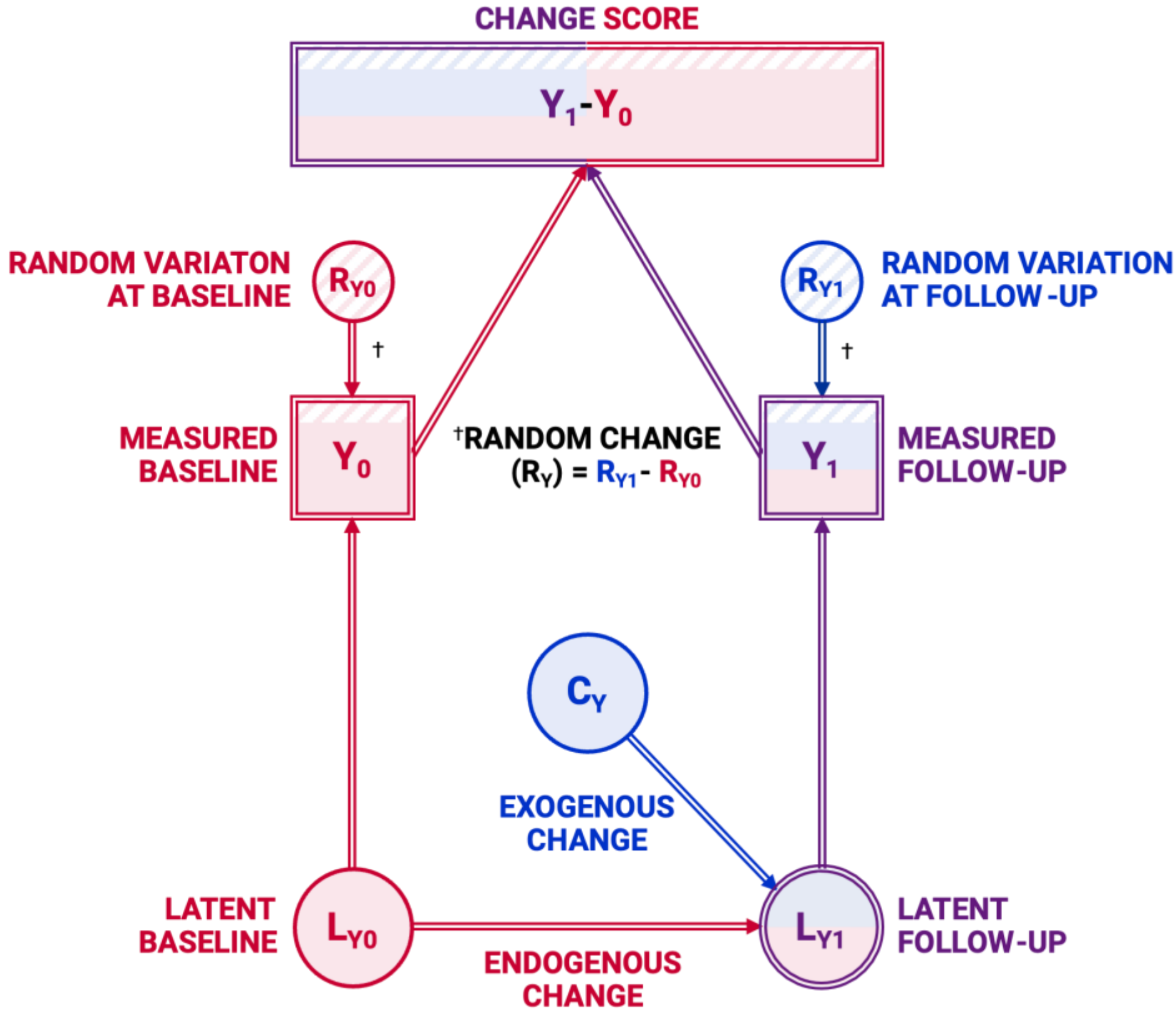

*$Y_0$ and $Y_1$ are the instantaneous, or 'measured', versions of the variable at baseline and follow-up respectively; $L_{Y0}$ and $L_{Y1}$ are the underlying latent, or 'true', versions of the variable at baseline and follow-up respectively; $R_{Y0}$ and $R_{Y1}$ represents all sources of transient random variation, such as transient classical measurement error, in the baseline and follow-up measure respectively; $C_Y$ represents all sources of exogenous change between the baseline and follow-up variable. The overall random change ($R_Y$) between the measured variables is determined by the difference in the random variation in the baseline measure ($R_{Y0}$) and follow-up measure ($R_{Y1}$), i.e., $R_Y = R_{Y1} - R_{Y0}$. Measurable variables are depicted as squares, latent variables are depicted as circles, and the derived change score variable is depicted as a rectangle spanning from the baseline measurement period to the follow-up measurement period to represent the fact that it crystallises jointly across both timepoints (since it contains information from both). Variables depicted with double outlines (e.g., $Y_1$) are fully determined by their ancestors and have no additional sources of variation.[34] The shading signifies the variance contributed by the different elements; solid red is the underlying component of the baseline measure, solid blue is the exogenous change, hashed red is the random variation in the baseline measure, and hashed blue is the random variation in the follow-up measure. The hashed red and hashed blue shading, representing $R_{Y0}$ and $R_{Y1}$ respectively, contribute equally to the derived change score variable since the overall random change ($R_Y$) is composed equally from $Y_0$ and $Y_1$.*

## DEFINING THE TARGET ESTIMAND

Of the three reasons that a variable may change value, only exogenous change is potentially modifiable by intervening on other baseline variables because all endogenous change has been pre-determined and all random change is beyond influence. A typical causal question concerning the effect of a baseline exposure (e.g., $X$) on the subsequent 'change' in an outcome between baseline (e.g., $Y_0$) and follow-up (e.g., $Y_1$) is therefore ultimately interested in the effect of that exposure on the exogenous change in the outcome (e.g., $C_Y$). This can be mapped onto two types of estimand, depending on the relationship between the exposure ($X$) and the baseline outcome ($Y_0$). When $X$ cannot cause $Y_0$, as in a randomised controlled trial or observational study where $Y_0$ precedes $X$, the effect of $X$ on $C_Y$ corresponds to the average causal effect of $X$ on $Y_1$, since $X$ causes $Y_1$ entirely through $C_Y$ (**Figure 3a**). Conversely, when $X$ causes $Y_0$, as in Lord's original example (**Figure 3b**), the effect of $X$ on $C_Y$ corresponds to the causal effect of $X$ on $Y_1$ acting through $C_Y$ and not $Y_0$, i.e., a controlled direct causal effect of $X$ on $Y_1$ for fixed levels of $Y_0$. In Lord's original example, there was no interaction between $X$ and $Y_0$, meaning the controlled direct effect of $X$ on $Y_1$ is invariant across values of $Y_0$ and coincides with the natural direct effect.[22] The estimand for Lord's original enquiry can therefore be expressed as the controlled direct effect of sex on follow-up weight at any fixed level of baseline weight.

## Figure 3

Directed acyclic graphs showing the two main ways that an exposure ($X$) could be related to an outcome at baseline ($Y_0$) and follow-up ($Y_1$) in observational data.

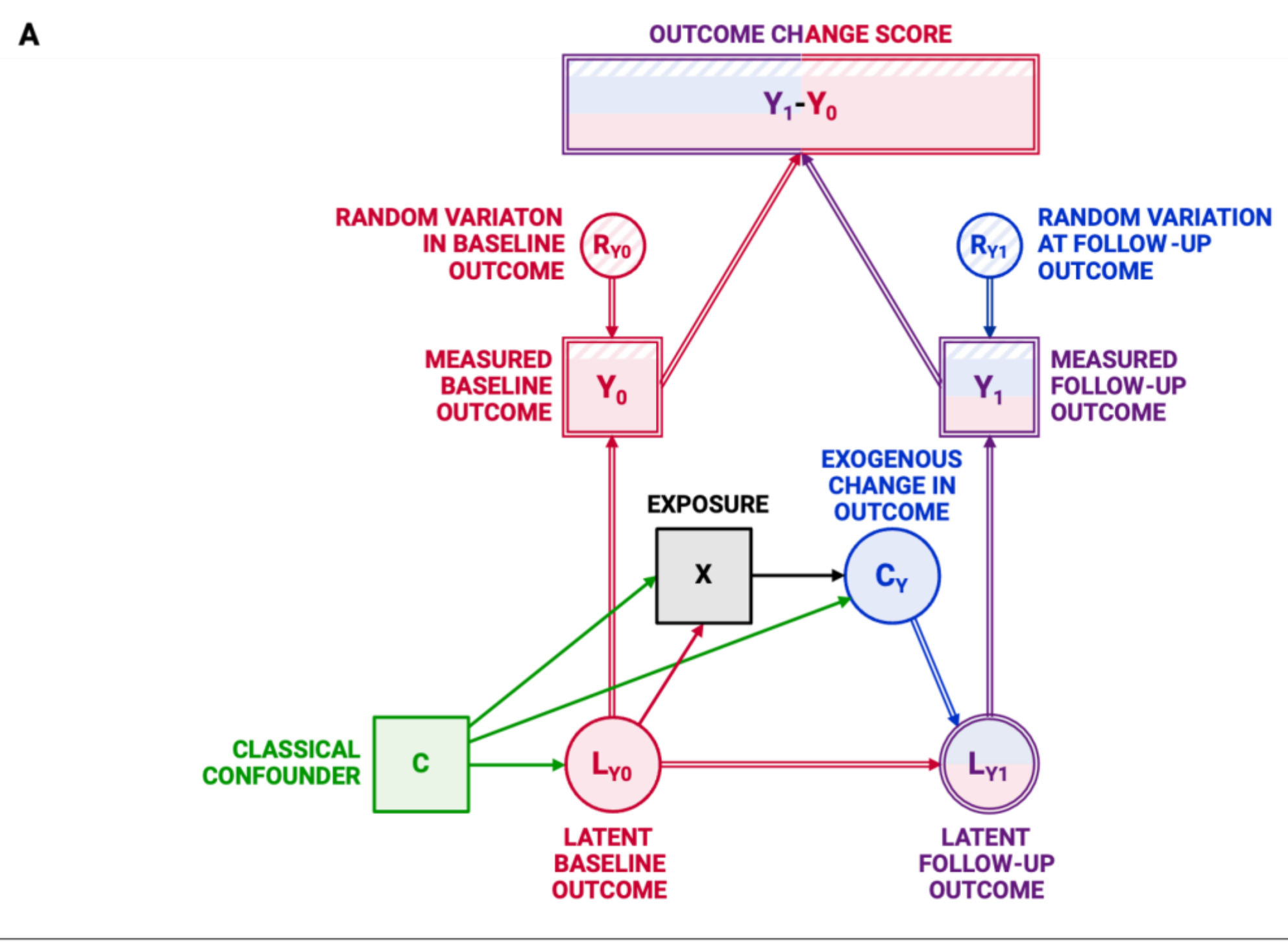

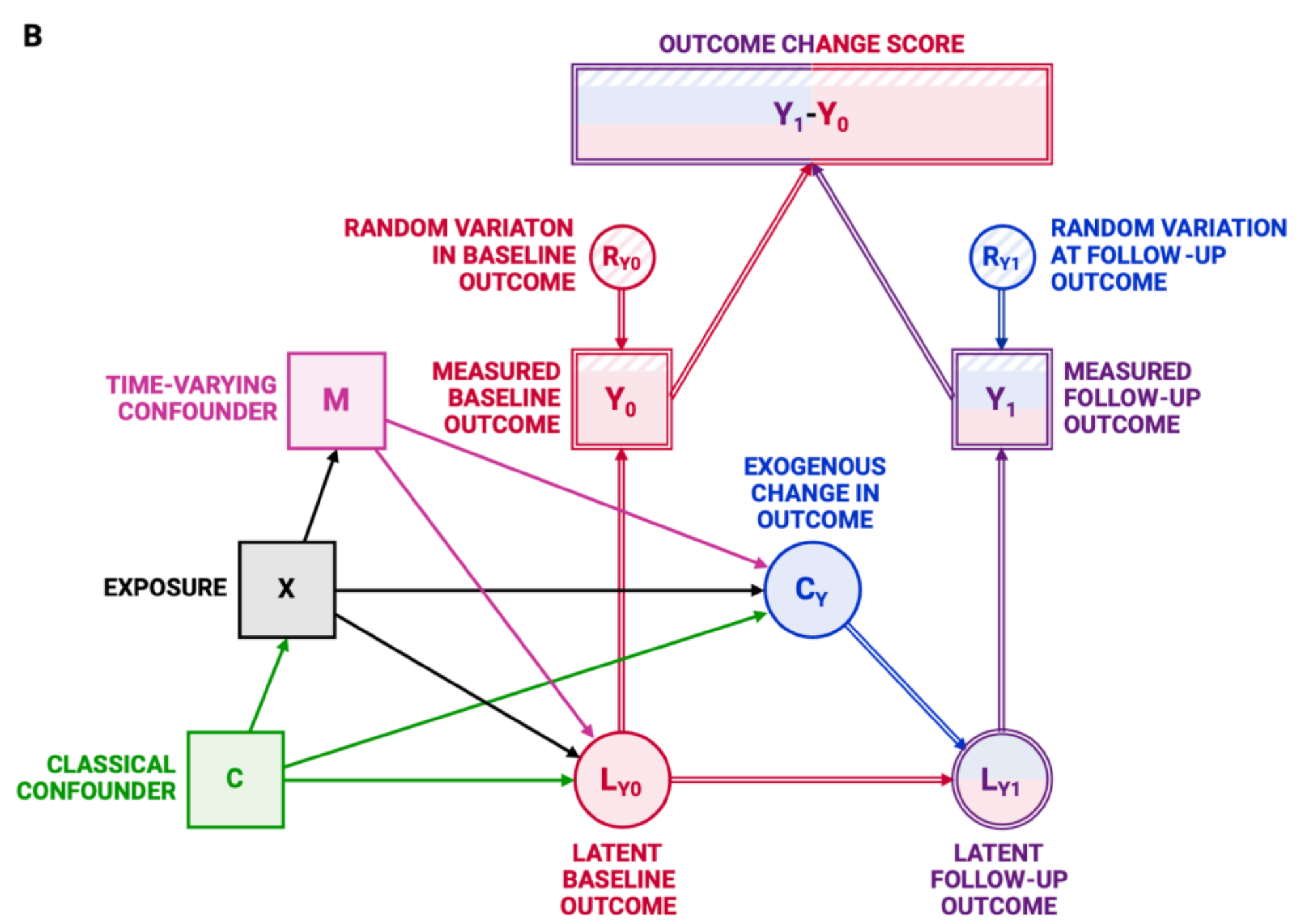

*In panel A, $Y_0$ occurs before $X$ and is a classical confounder for the effect of $X$ on $Y_1$. In panel B, $Y_0$ occurs after $X$ and mediates part of the effect of $X$ on $Y_1$. In both cases, $X$, $Y_0$, $Y_1$ may also be associated due to common causes (i.e., classical confounder). In the mediation scenario depicted in panel B, $X$ may also be related to $Y_0$ and $Y_1$ via a time-varying confounder ($M$), as shown.*

*Outcome variables are depicted as described in Figure 2, with instantaneous, or 'measured', versions ($Y_0$ and $Y_1$) caused by a combination of their underlying latent, or 'true' versions ($L_{Y0}$ and $L_{Y1}$) and their transient random variation, such as transient classical measurement error ($R_{Y0}$ and $R_{Y1}$).*

## SAME QUESTION, DIFFERENT BIASES: THE SOLUTION TO LORD'S PARADOX

The solution to Lord's 'paradox' is not, in fact, that the two approaches are asking different questions. Both approaches can theoretically be used to estimate the same estimand, but only under certain assumptions. When these assumptions are not met, as in Lord's original example, the two approaches will typically produce estimates that are biased in opposite directions. To explain, consider a perfectly measured follow-up instance of a variable ($Y_1$), that is determined by a function of: 1) all endogenous change from the baseline instance of that variable ($Y_0$) and 2) all exogenous change ($C_Y$) such that $E[Y_1] = f_1(Y_0) + f_2(C_Y)$. Although exogenous change cannot be directly measured, it can, in theory, be estimated from the difference between the observed follow-up measure and the value *expected* due to the endogenous change from baseline, i.e., $E[C_Y] = Y_1 - f(Y_0)$. A simple ANCOVA analysis attempts to do this by modelling $f(Y_0)$ as a linear function, e.g., $E[C_Y] = Y_1 - \beta_1 Y_0$. Unfortunately, this approach is susceptible to bias whenever $f(Y_0)$ is nonlinear and/or $Y_0$ is imperfectly measured. Since $Y_0$ will often be imperfectly measured, this means the follow-up conditional on baseline approach is frequently prone to bias. Typically, the impact is an *inflated* estimate of $E[C_Y]$, because measurement error in $Y_0$ tends to dilute the estimate of $\beta_1$ (via regression dilution)[23], meaning a diluted version of $\beta_1 Y_0$ gets subtracted from $Y_1$.[23]

In contrast, the change score approach is prone to making inflated subtractions from $Y_1$, leading to *diluted* or *opposite-sign* estimates of $E[C_Y]$. To understand, consider the ANCOVA function proposed above which we can rewrite as: $Y_1 - \beta_1 Y_0 = E[C_Y]$ . This can be converted into a change score by subtracting $[1 - \beta_1]Y_0$ from both sides to produce: $[Y_1 - Y_0] = E[C_Y] - [1 - \beta_1]Y_0$. Rather than isolating exogenous change, a change score hence contains an additional $-[1 - \beta_1]Y_0$ subtraction term.

## THE IMPACT OF THE CHANGE SCORE SUBTRACTION TERM

The extent to which the $-[1 - \beta_1]Y_0$ subtraction term introduces bias in analyses of change scores depends on several factors, most importantly the relationship between the exposure and the baseline outcome.

In a randomised experiment or natural experiment, where the exposure is exogenous, analyses of change scores will provide accurate estimates of the causal effect of the exposure on exogenous change in the outcome because there is no association between the exposure and the baseline outcome, meaning the $-[1 - \beta_1]Y_0$ term has no impact besides increasing the variance compared with analysing the follow-up outcome directly.[15]

In observational data, however, any association between the exposure and the baseline outcome will introduce bias via the $-[1 - \beta_1]Y_0$ term unless two additional assumptions hold.[24,25] The first is the parallel trends assumption, which requires that any differences in the outcome between exposure groups should be stable over time, excluding any effect of the exposure itself.[24,25] This essentially requires that any association between the exposure and baseline outcome is entirely due to time-fixed confounders with constant effects on the outcome over time.[24,25] Consequently, if the exposure either causes, or is caused by, the baseline outcome, as in Lord's original example, the parallel trends assumption will not hold and bias is expected.[26] The second is the no anticipation assumption, which requires that the exposure must not have any effect on the outcome prior to baseline.[24] This is clearly violated in Lord's original scenario, and indeed any scenario where the baseline outcome is a mediator, because the exposure causes the outcome before baseline.[24]

When these assumptions are not met, the size of the bias created by the $-[1-\beta_1]Y_0$ subtraction term will primarily depend on the size of the association between the exposure and the baseline outcome. In Lord's original example, that association was *exactly* equal to the average causal effect of the exposure on the follow-up outcome, leading to their perfect cancelation and hence the first statistician's conclusion of 'no overall change'. Where the association is larger, it can overpower the causal effect of the exposure on the follow-up outcome, resulting in coefficients with opposite signs to the average or direct causal effects of interest.[9]

## FURTHER PROBLEMS AND SOLUTIONS IN FOLLOW-UP CONDITIONAL ON BASELINE ANALYSES

Analyses of follow-up conditional on baseline also require several assumptions to be satisfied to return accurate estimates in observational data. The first of these assumptions, as previously discussed, is that the baseline outcome needs to be perfectly measured. When, instead, the baseline outcome is imperfectly measured, it will be imperfectly conditioned, leading to incomplete closure of the nuisance causal path via the baseline outcome. This typically results in an inflated effect estimate.[23] While such bias is essentially ubiquitous, it can be reduced by using the best available measurement method and/or collecting and utilizing multiple measures of the baseline outcome.[27]

The second assumption is that the follow-up conditional on baseline model must correctly parameterise the relationship between the exposure and the baseline outcome to completely close the nuisance path via the baseline outcome. In Lord's original example, this relationship was linear, as shown by the straight parallel lines. This allowed the effect to be robustly estimated with a simple additive linear (i.e. ANCOVA) model. However, wherever the relationship is nonlinear, it would need to be correctly modelled to provide an accurate estimate.[18]

The third assumption is that all other sources of confounding must be accounted for. The simple follow-up conditional on baseline model shown in **Equation 2** offers no protection against bias from other sources of classical confounding besides the baseline outcome. In Lord's original example, the lack of adjustment for classical confounders (i.e., mutual causes of the exposure and outcome) is probably reasonable, since the exposure (sex) is exogenous. However, most similar scenarios involving analyses of change would need to condition on other causes of the exposure and follow-up outcome.[28] Mediator-outcome confounding and time-varying confounding must also be considered when the baseline outcome is a mediator.[29] In Lord's thought experiment, there is no mention of mediator-outcome confounders (i.e., mutual causes of the outcome at baseline and follow-up) or time-varying confounders (i.e., mediator-outcome confounders caused by the exposure). However, if the study was repeated in real data, many such variables would probably exist. Physical activity levels, for example, would likely be influenced by sex and in turn contribute to weight at the start and end of the year. Under such circumstances, neither the simple model shown in **Equation 2** nor any similar approach based on conditioning can produce accurate direct causal effect estimates. Appropriate multivariate methods, such as g-methods, would be required instead.[29]

Finally, the positivity assumption must also be satisfied, i.e., there must be sufficient overlap of the baseline outcome distribution between exposure groups. In Lord's original example, this means that each baseline weight value should be possible among both boys and girls.[30] Unfortunately, the data clouds in Lord's original figure, show very limited overlap in baseline weight between the sexes (**Figure 1**). In such situations, the accuracy of the resulting estimate becomes highly model-dependent.[31]

Although this may be reasonable in Lord's original scenario, where we know that all relationships are additive and linear, it is generally advisable to restrict such analyses (and accordingly redefine the target estimands) to the subsample of participants whose baseline outcome values fall within the overlapping region.[31]

## ILLUSTRATING LORD'S 'PARADOX' WITH SIMULATED DATA

To help demonstrate Lord's 'paradox', we simulated the archetypical scenario of an outcome measured in two groups at two time points, where the two groups differ by the same amount at baseline and follow-up.[32] We could give these variables any name and make them as well-, or poorly-, defined as we want, but for simplicity we called them sex, baseline weight, and follow-up weight, as in Lord's original example. We made sex cause underlying weight at baseline and follow-up, meaning sex causes exogenous change in weight. We also simulated transient random variation in the measured versions of baseline and follow-up weight, which may be interpreted as transient classical measurement error. We also simulated a second measure of baseline weight, with its own measurement error, to demonstrate the potential value of taking repeated baseline measures. In a second simulation, we introduced an additional time-varying confounder, physical activity, that is caused by sex, and in turn causes both baseline and follow-up weight. Further details of the simulation are available in the **Supplementary materials** and the simulation code is available on Github at: **https://github.com/pwgtennant/lords-paradox**.

**Figure 4** and **Supplementary figure S2** presents the simulated data in the style of Lord's original diagram, for the scenarios without and with time-varying confounding, respectively. **Table 1** reports the estimands and results in both scenarios for various analytical approaches using both the 'measured' and 'true' versions of the outcome. In all scenarios, the change score analyses produced null coefficients that misleadingly imply that the exposure does *not* cause 'change', while the follow-up conditional on baseline analyses produced positive coefficients that correctly imply that the exposure does cause 'change'. However, the follow-up conditional on baseline analyses only produced unbiased estimates in the absence of both time-varying confounding and measurement error in the baseline outcome. Time-varying confounding by physical activity led to diluted estimates, measurement error in baseline weight led to inflated estimates, and measurement error in follow-up weight led to increased uncertainty (signified by wider simulation intervals). Appropriately modelling physical activity using g-computation recovered accurate estimates in the presence of time-varying confounding, but only in the absence of measurement error in the baseline outcome. Conditioning on an average of two baseline outcome measures produced more accurate estimates than conditioning on a single baseline outcome measure, although some bias remained. Analyses of change scores that also conditioned on the baseline outcome returned identical estimates as analyses of follow-up conditional on baseline, showing these approaches are equivalent.

## Figure 4

Scatterplot of the simulated data for the scenario without time-varying confounding in the style of Lord's original diagram (as reproduced in **Figure 1**), showing the 'measured' follow-up weight ($Y_1$) plotted against the 'measured' baseline weight ($Y_0$), for boys and girls separately.

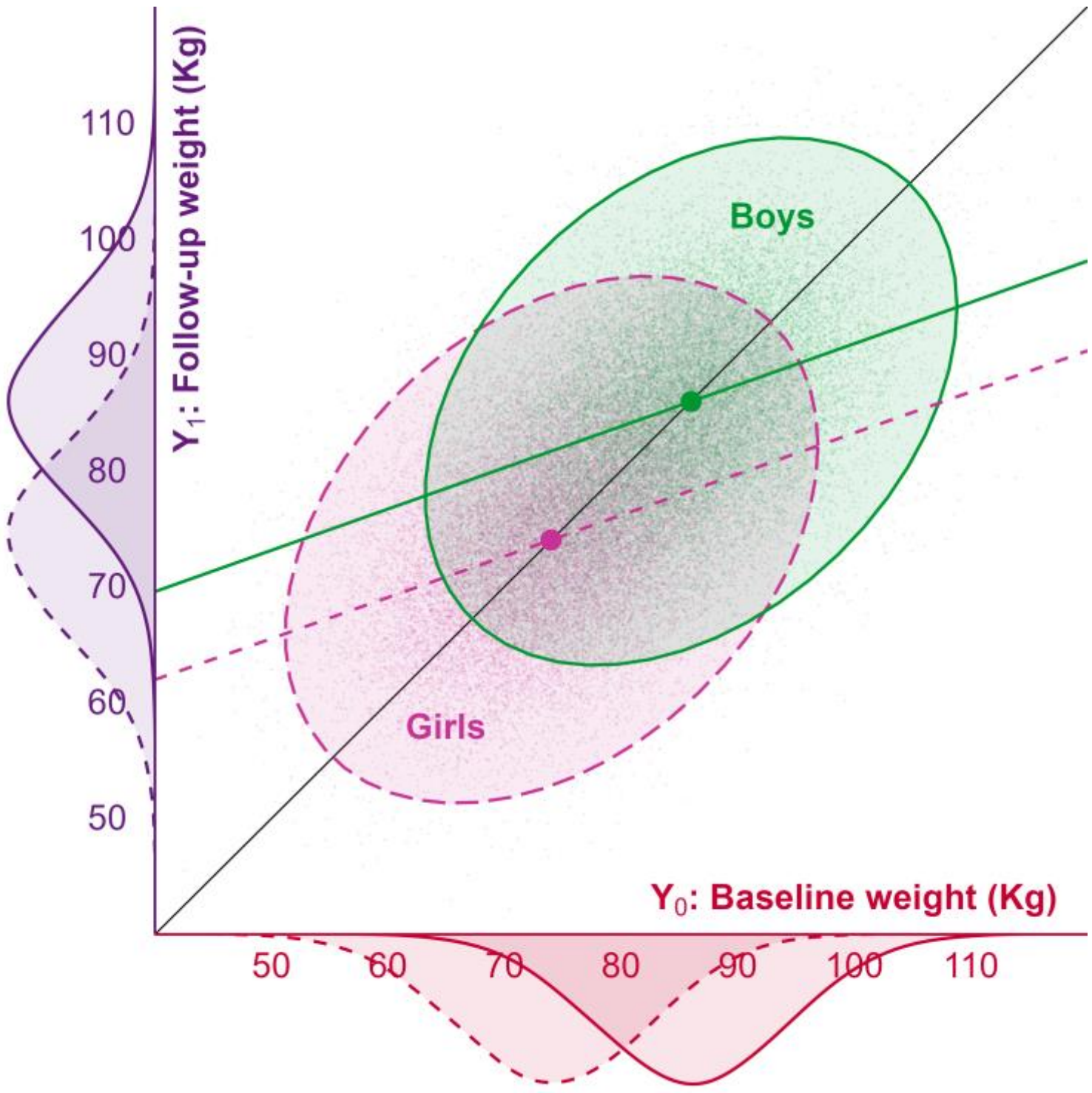

*The ellipses enclose 99.5% of the scatter points for the boys (depicted in green with solid lines) and the girls (depicted in pink with dashed lines). Smoothed kernel density plots of the initial weights ($Y_0$), shown in red, with solid lines for boys and dashed lines for girls) are indicated at the bottom and smoothed kernel density plots of final weight ($Y_1$), shown in purple, again with solid lines for boys and dashed lines for girls) are indicated on the left. Separate linear regression lines for the relationship between the baseline and follow-up weight are depicted for boys (solid green) and girls (dashed pink). Further details of the simulation are available in the* ***Supplementary materials*** *and the simulation code is available on Github at:* ***https://github.com/pwgtennant/lords-paradox****. A similar plot for the scenario with time-varying confounding is presented in* ***Supplementary figure S2****.*

## Table 1

Estimates produced by four possible analytical approaches to exploring the controlled direct effect of sex on follow-up weight at any fixed level of baseline weight*, without and with transient classical measure error and time-varying confounding from physical activity.

| **Approach** | **Estimate (95% SI)** | | | | | | | | | | |
|---|---|---|---|---|---|---|---|---|---|---|---|
| | Without time-varying confounding | | | | | | With time-varying confounding | | | | |
| | True follow-up | | | Measured follow-up | | | True follow-up | | | Measured fol | |
| | True baseline | Single - measured baseline | Double- measured baseline | True baseline | Single - measured baseline | Double- measured baseline | True baseline | Single - measured baseline | Double- measured baseline | True baseline | Single measur baselin |
| 1) **Approach 1 (change score)**: Analysis of the outcome change score by linear regression | 0.0kg (-0.8, 0.8) | 0.0kg (-1.0, 1.0) | 0.0kg (-0.9, 0.9) | 0.0kg (-1.0, 1.0) | 0.0kg (-1.1, 1.1) | 0.0kg (-1.0, 1.1) | 0.0kg (-0.5, 0.5) | 0.0kg (-0.8, 0.8) | 0.0kg (-0.7, 0.7) | 0.0kg (-0.8, 0.8) | 0.0kg (-0.9, 0. |
| 2) **Approach 2 (follow-up conditional on baseline)**: Analysis of the follow-up outcome adjusted for the baseline outcome by linear regression | **6.0kg (5.0, 7.0)** | 7.8kg (6.8,8.7) | 7.0kg (6.1,8.0) | **6.0kg (4.8, 7.2)** | 7.8kg (6.6, 8.9) | 7.0kg (5.9, 8.3) | 2.4kg (1.7, 3.1) | 5.3kg (4.5, 6.0) | 4.1kg (3.3, 4.8) | 2.4kg (1.4, 3.4) | 5.3kg (4.3, 6. |
| 3) **Approach 3 (g-computation)**: Analysis of the joint effects of the exposure and baseline outcome accounting for the time-varying confounder[a] | N/A | N/A | N/A | N/A | N/A | N/A | **6.0kg (5.2, 6.9)** | 9.1kg (8.2, 10.0) | 8.1kg (7.2, 9.0) | **6.0kg (4.7, 7.3)** | 9.1kg (7.9, 10. |
| 4) **Approach 4 (adjusted change score)**: Analysis of the outcome change score adjusted for the baseline outcome by linear regression | **6.0kg (5.0, 7.0)** | 7.8kg (6.8,8.7) | 7.0kg (6.1,8.0) | **6.0kg (4.8, 7.2)** | 7.8kg (6.6, 8.9) | 7.0kg (5.9, 8.3) | 2.4kg (1.7, 3.1) | 5.3kg (4.5, 6.0) | 4.1kg (3.3, 4.8) | 2.4kg (1.4, 3.4) | 5.3kg (4.3, 6. |

*$X$ = exposure (sex), $Y_0$ = mediator (baseline weight), $Y_1$ = outcome (follow-up weight). The ground truth controlled direct effect in all scenarios is 6.0kg. Results shown in bold represent scenarios where the ground truth effect has been estimated without bias. *In Lord's original scenario, and hence in our simulation, there is no interaction between the exposure and mediator; the controlled direct effect is therefore identical for all values of $Y_0$, and also equals the natural direct effect.[22] In the presence of nonlinearities and interactions, the specific controlled direct effect estimated would depend on the selected (or average) value of the mediator. See the **Supplementary materials** for more details of the data simulations.*

## CONCLUSION

Lord 'paradox' has attracted considerable debate for several decades.[1–7] This may partly stem from the ambiguity of Lord's original description, which provides endless scope for reinterpretation. Yet even when the context is stripped away, and the variables are relabelled $X$, $Y_0$, and $Y_1$, a genuine puzzle remains. In observational data, why do two ostensibly reasonable approaches to estimating the effect of an exposure on 'change' in an outcome sometimes return radically different results? The answer is surprisingly simple. Each approach is prone to bias, and these biases typically act in opposite directions. For analyses of change scores, this bias can be particularly severe when the main assumptions of the approach are not met. These will fail whenever the exposure causes, or is caused by, the baseline outcome, as in Lord's original example and in most observational data. Analyses of change scores are therefore best reserved for natural experiment scenarios where the exposure is plausibly exogenous.[20,33]

## REFERENCES

1. Lord FM. A paradox in the interpretation of group comparisons. *Psychological Bulletin*. 1967;**68**(5):304–305.
2. Locascio JJ, Cordray DS. A Re-Analysis of 'Lord's Paradox'. *Educational and Psychological Measurement*. 1983;**43**(1):115–126.
3. Holland PW, Rubin DB. Research Designs and Causal Inferences: On Lord's Paradox. In: Pearson RW, Boruch RF, editors. *Survey Research Designs: Towards a Better Understanding of Their Costs and Benefits: Prepared under the Auspices of the Working Group on the Comparative Evaluation of Longitudinal Surveys Social Science Research Council*. New York, NY: Springer US; 1986. p. 7–37.
4. Lund T. Lord's Paradox Re-examined. *Scandinavian Journal of Educational Research*. 1999;**43**(1):41–55.
5. Pearl J. Lord's Paradox Revisited – (Oh Lord! Kumbaya!). *Journal of Causal Inference*. 2016;**4**(2):20160021.
6. Xiao Z, Higgins S, Kasim A. An Empirical Unraveling of Lord's Paradox. *The Journal of Experimental Education.* 2017;**87**(1):17-32.
7. Wijayatunga P. Resolving the Lord's Paradox. ArXiv; 2018. DOI: arXiv:1804.07923
8. Lord FM. Statistical adjustments when comparing preexisting groups. *Psychological Bulletin*. 1969 Nov;**72**(5):336–337.
9. Wright DB. Allocation to groups: Examples of Lord's paradox. *British Journal of Educational Psychology*. 2020;**90**(S1):35–49.
10. London K, Wright DB. Analyzing change between two or more groups: Analysis of variance versus analysis of covariance. *Handbook of developmental research methods*. New York, NY, US: The Guilford Press; 2012. p. 279–290.
11. Hand DJ. Deconstructing Statistical Questions. *Journal of the Royal Statistical Society. Series A (Statistics in Society).* 1994;**157**(3):317.
12. Glymour MM, Spiegelman D. Evaluating Public Health Interventions: 5. Causal Inference in Public Health Research—Do Sex, Race, and Biological Factors Cause Health Outcomes? *American Journal of Public Health*. 2017;**107**(1):81–85.

13. Senn S. Rothamsted Statistics meets Lord's Paradox. https://errorstatistics.com/2018/11/11/stephen-senn-rothamsted-statistics-meets-lords-paradox-guest-post/. Accessed January 19, 2026.

14. VanderWeele TJ. On Well-defined Hypothetical Interventions in the Potential Outcomes Framework. *Epidemiology*. 2018;**29**(4):e24–e25.

15. Vickers AJ, Altman DG. Statistics Notes: Analysing controlled trials with baseline and follow up measurements. *BMJ*. 2001;**323**(7321):1123–1124.

16. Van Breukelen GJP. ANCOVA versus change from baseline had more power in randomized studies and more bias in nonrandomized studies. *Journal of Clinical Epidemiology*. 2006;**59**(9):920–925.

17. Glymour MM, Weuve J, Berkman LF, Kawachi I, Robins JM. When Is Baseline Adjustment Useful in Analyses of Change? An Example with Education and Cognitive Change. *American Journal of Epidemiology*. 2005;**162**(3):267–278.

18. Senn S. Change from baseline and analysis of covariance revisited. *Statistics in Medicine*. 2006;**25**(24):4334–4344.

19. Shahar E, Shahar DJ. Causal diagrams and change variables. *Journal of Evaluation in Clinical Practice*. 2012;**18**(1):143–148.

20. Tennant PWG, Arnold KF, Ellison GTH, Gilthorpe MS. Analyses of 'change scores' do not estimate causal effects in observational data. *International Journal of Epidemiol*. 2022;**51**(5):1604–1615.

21. Hernan MA, Cole SR. Invited Commentary: Causal Diagrams and Measurement Bias. *American Journal of Epidemiology*. 2009;**170**(8):959–962.

22. VanderWeele TJ. Mediation Analysis: A Practitioner's Guide. *Annual Reviews of Public Health*. 2016;**37**(1):17–32.

23. Keogh RH, Shaw PA, Gustafson P, et al. STRATOS guidance document on measurement error and misclassification of variables in observational epidemiology: Part 1—Basic theory and simple methods of adjustment. *Statistics in Medicine*. 2020;**39**(16):2197–2231.

24. Roth J, Sant'Anna PHC, Bilinski A, Poe J. What's trending in difference-in-differences? A synthesis of the recent econometrics literature. *Journal of Econometrics*. 2023;**235**(2):2218–2244.

25. Caniglia EC, Murray EJ. Difference-in-Difference in the Time of Cholera: a Gentle Introduction for Epidemiologists. *Current Epidemiology Reports*. 2020;**7**(4):203–211.

26. Renson A, Dukes O, Shahn Z. Using causal diagrams to assess parallel trends in difference-in-differences studies. ArXiv; 2025. DOI: arXiv:2505.03526

27. Blackwell M, Honaker J, King G. A Unified Approach to Measurement Error and Missing Data: Details and Extensions. *Sociologic Methods & Research*. 2017;**46**(3):342–369.

28. Tennant PWG, Murray EJ, Arnold KF, et al. Use of directed acyclic graphs (DAGs) to identify confounders in applied health research: review and recommendations. *International Journal of Epidemiology*. 2021;**50**(2):620–632.

29. Mansournia MA, Etminan M, Danaei G, Kaufman JS, Collins G. Handling time varying confounding in observational research. *BMJ*. 2017 Oct 16;j4587.

30. Westreich D, Cole SR. Invited Commentary: Positivity in Practice. *American Journal of Epidemiology*. 2010;**171**(6):674–677.

31. King G, Zeng L. The Dangers of Extreme Counterfactuals. *Political Analysis*. 2006;**14**(2):131–159.

32. Al Hajj GS, Pensar J, Sandve GK. DagSim: Combining DAG-based model structure with unconstrained data types and relations for flexible, transparent, and modularized data simulation. Adabor ES, editor. *PLOS One*. 2023;**18**(4):e0284443.

33. Cunningham S. Causal inference: the mixtape. New Haven London: Yale University Press; 2021.

34. Berrie L, Arnold KF, Tomova GD, Gilthorpe MS, Tennant PWG. Depicting deterministic variables within directed acyclic graphs: an aid for identifying and interpreting causal effects involving derived variables and compositional data. *American Journal of Epidemiology*. 2025;**194**(2):469–479.

## SUPPLEMENTARY MATERIALS

### Supplementary methods

We conducted two simulations for two scenarios that both reproduce the data patterns and paradox that was first described by Lord in 1967. The data generating mechanisms behind these two scenarios are depicted in **Supplementary Figure S1**. The first scenario (Scenario 1) is free from mediator-outcome confounding and the second scenario (Scenario 2) is complicated by mediator-outcome confounding.

Data were simulated using the `DagSim` 1.0.10 package for Python 3.11.[32] `DagSim` for Python is a flexible framework for simulating data from a DAG that allows for a variety of variable types, distributions, and relationships.

In `DagSim`, simulations are performed iteratively based on the topological (i.e., time) order of the variables. For each simulation, we first supplied `DagSim` with the relevant data generating DAG. We then defined functions for each variable within the DAG, based on the relationship with each parent node and adding any necessary noise. We aimed to create baseline weight and follow-up weight variables with means of 80kg and standard deviations (SDs) of 9kg for the underlying latent, or 'true', versions and 10kg for the instantaneous, or 'measured' versions. We aimed for a weight difference of 12kg between boys and girls at both time points.

In scenario 1, the exposure, sex ($X$), was defined first from a symmetrical binomial distribution, with girls = -1 and boys = 1 to provide a mean of 0 and SD of 1. Next, the underlying latent, or 'true', version of baseline weight ($L_{Y0}$) was defined from sex, with an unstandardized path coefficient ($\beta_1$) of 6 (i.e., boys being heavier than girls), and the necessary (normally distributed) noise to return the target SD of 9kg. To create the instantaneous, or 'measured', version of baseline weight ($Y_0$) an additional (normally distributed) noise variable ($R_{Y0}$) was simulated to represent all transient random sources of variation at baseline (such as measurement error). The instantaneous, or 'measured', version of baseline weight was then defined from the 'true' baseline weight, with an unstandardized path coefficient ($\beta_2$) of 1, and the random variation at baseline, with an unstandardized path coefficient ($\beta_3$) of $\sqrt{(10^2 - 9^2)/9^2} \approx 0.484$. $\beta_3$ was determined by rearranging the equation for the total variance of 'true' follow-up weight ($Y_0$) setting a target SD of 10kg, i.e.,

$$Var(Y_0) = Var(L_{Y0}) + Var(R_{Y0}) + 2Cov(L_{Y0}, R_{Y0}) \Rightarrow$$

$$10^2 = (9 \times \beta_2)^2 + (9 \times \beta_3)^2 + 0 \Rightarrow$$

$$100 = 81 + 81 \times {\beta_3}^2 + 0 \Rightarrow$$

$$\beta_3 = \sqrt{19/81} \approx 0.484.$$

To show the benefits of collecting and utilizing multiple measures of the baseline outcome, a second 'measured' version of baseline weight ($Y_{0_2}$) was also created in the same way from the true baseline weight and a second noise variable ($R_{Y0_2}$), and an average variable was created ($Y_{0_avg} = [Y_0 + Y_{0_2}]/2$). These are omitted from **Supplementary figure S1** for simplicity.

Next, exogenous change ($C_Y$) was defined from sex with an unstandardized path coefficient ($\beta_4$) of $6/\sqrt{2.1\dot{1}} \approx 4.129$ (i.e., boys having more exogenous change than girls). $\beta_4$ was determined together with the path coefficient for the effect of exogenous change on the latent, or 'true', version of follow-up weight using the approach discussed below. They were chosen to ensure that the controlled direct causal effect of sex on follow-up weight (6kg) was equal to the mediated causal effect of sex that acts through baseline weight (6kg); this is necessary to ensure the 'no overall change' observed in Lord's original thought experiment.

The latent, or 'true', version of follow-up weight ($L_{Y0}$) was defined from 'true' baseline weight, with an unstandardized path coefficient ($\beta_5$) of 0.5 (i.e., baseline weight causing follow-up weight), and exogenous change, with an unstandardized path coefficient ($\beta_6$) of $0.5\sqrt{2.1\dot{1}} \approx 0.726$ (i.e., exogenous change causing follow-up weight). $\beta_4$ and $\beta_6$ were jointly determined by rearranging the equation for the total variance of the 'true' follow-up weight ($L_{Y1}$), setting a target SD of 9kg, i.e.,

$$Var(L_{Y1}) = Var(L_{Y0}) + Var(C_Y) + 2Cov(L_{Y0}, C_Y) \Rightarrow$$

$$9^2 = (9 \times \beta_5)^2 + (9 \times \beta_6)^2 + 2 \times \beta_5 \times \beta_6 \times \beta_1 \times \beta_4 \Rightarrow$$

$$81 = 81 \times 0.5^2 + 81 \times 0.5x^2 + 2 \times 0.5 \times 0.5x \times 6 \times 6/x \Rightarrow$$

$x = \sqrt{2.1\dot{1}}$ and therefore $\beta_4 = 6/\sqrt{2.1\dot{1}} \approx 4.129$, and $\beta_6 = 0.5\sqrt{2.1\dot{1}} \approx 0.726$

Since the second statistician in Lord's original example found that "*the slope of the regression line of final weight on initial weight is essentially the same for the two sexes",* we did not introduce an interaction effect between the exposure (sex) and mediator (baseline weight) on the 'true' outcome (follow-up weight); the consequence of this is that an accurate estimate of the controlled direct effect does not require an interaction term, and the controlled direct effect is invariant across all values of the mediator (baseline weight) and coincides with the natural direct effect.[22]

The instantaneous, or 'measured', version of follow-up weight ($Y_1$) was defined from the 'true' follow-up weight, with an unstandardized path coefficient of 1, and the random variation at follow-up ($R_{Y1}$), with path coefficient of $\sqrt{(10^2 - 9^2)/9^2} \approx 0.484$. As above, the ≈0.484 path coefficient was chosen so that the target SD for the 'measured' version of follow-up weight was 10kg.

Finally, six change score variables were defined from the various different versions of baseline and follow-up weight, one subtracting 'true' baseline weight from 'true' follow-up weight ($L_{Y1} - L_{Y0}$), one subtracting the single 'measured' baseline weight from 'measured' follow-up weight ($Y_1 - Y_0$), one subtracting the average of the two 'measured' baseline weights from 'measured' follow-up weight ($Y_1 - Y_{0_avg}$), one subtracting 'true' baseline weight from 'measured' follow-up weight ($Y_1 - L_{Y0}$), one subtracting the single 'measured' baseline weight from the 'true' follow-up weight ($L_{Y1} - Y_0$), and one subtracting an average of the two 'measured' baseline weights from the 'true' follow-up weight ($L_{Y1} - Y_{0_avg}$). For simplicity, only one of these change score variables, the one made from differencing the single 'measured' versions ($Y_1 - Y_0$) is shown in **Supplementary figure S1**.

A similar procedure was followed to simulate the data for scenario 2, with the addition of a time-varying confounder, physical activity ($M$), which had a target mean of 30 minutes and was defined from sex, with path coefficient 5 (i.e., boys doing more physical activity than girls), and the necessary (normally distributed) noise to return a target SD of 10 minutes. In turn, physical activity was made to cause 'true' baseline weight, with an unstandardized path coefficient of -0.60 (i.e., each additional minute leading to a 0.60kg reduction in weight), and exogenous change in weight, with an unstandardized path coefficient of $-0.60/\sqrt{1.4\dot{4}} \approx -0.499$ (i.e., physical activity causing lower exogenous change in weight). To maintain the same total causal effects and to obtain the correct SDs, the unstandardized path coefficients between sex and exogenous change, between 'true' baseline weight and exogenous change, and between exogenous change and 'true' follow-up weight were all changed respectively to 9, $9/\sqrt{1.4\dot{4}} \approx 7.488$, and $0.5\sqrt{1.4\dot{4}} \approx 0.601$.

For both scenarios, we simulated 10,000 datasets each containing 1,000 observations. In both scenarios, the ground-truth controlled direct effect of sex (and hall/diet, since the two are inseparable) on 'true' follow-up weight, independent of baseline weight, was 6kg.

The four modelling approaches described in **Table 1** were run in each of the 10,000 datasets with the median coefficient and the 2.5th and 97.5th centiles (representing 95% simulation intervals, SIs) being presented in the table. G-computation models were conducted using the `CMAverse` package for R 4.5.1.[35] Figure 4 and Supplementary Figure 1 were created within single datasets that were boosted to have a sample of 10,000 observations, using the `ggplot2` package for R 4.5.1;[36] the ellipses were created using the `stat_ellipse` function, which uses a modified version of the method outlined by Fox and Weisberg (2011).[37]

## Supplementary figure S1

Directed acyclic graphs showing the data generating process used for the two simulations.

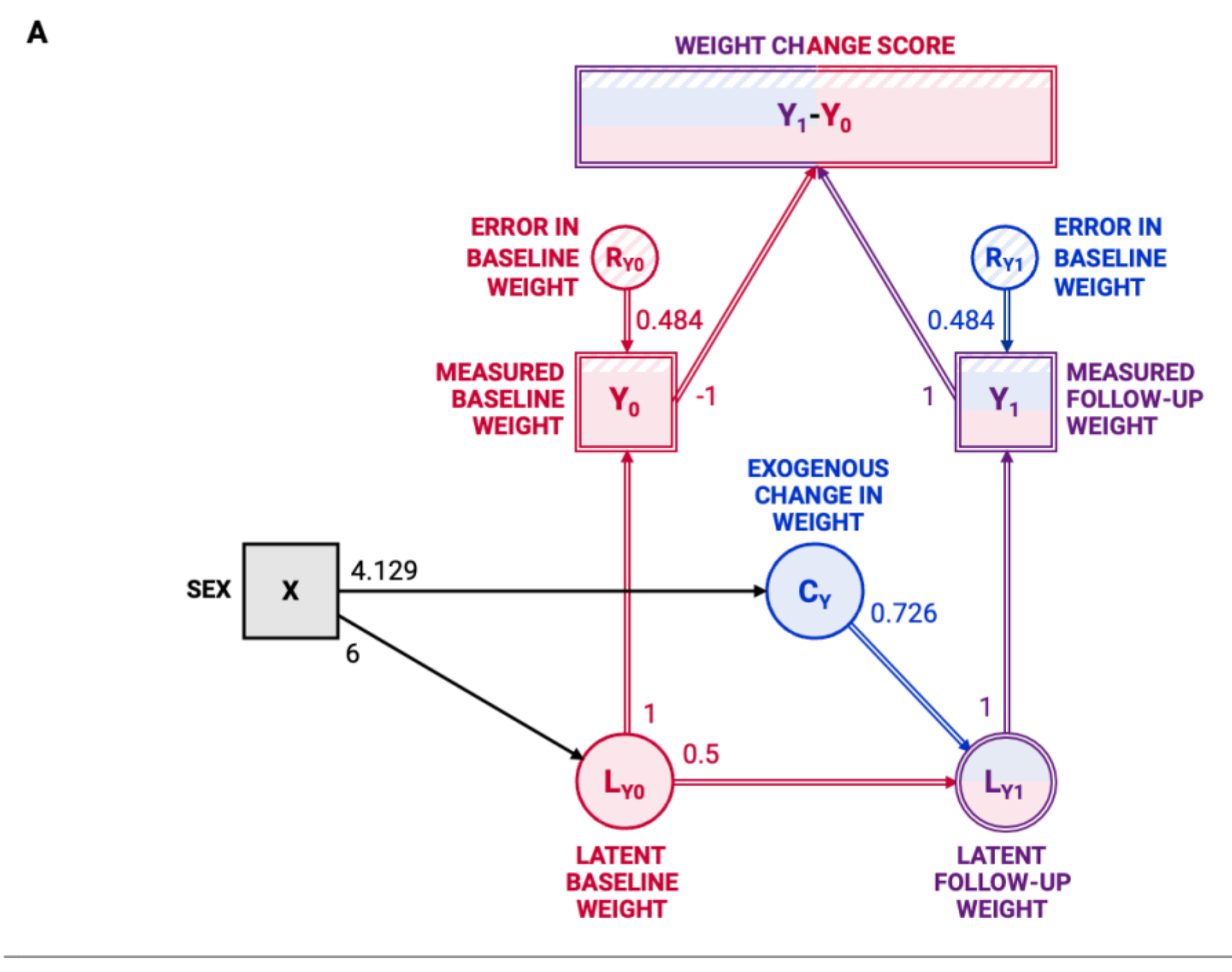

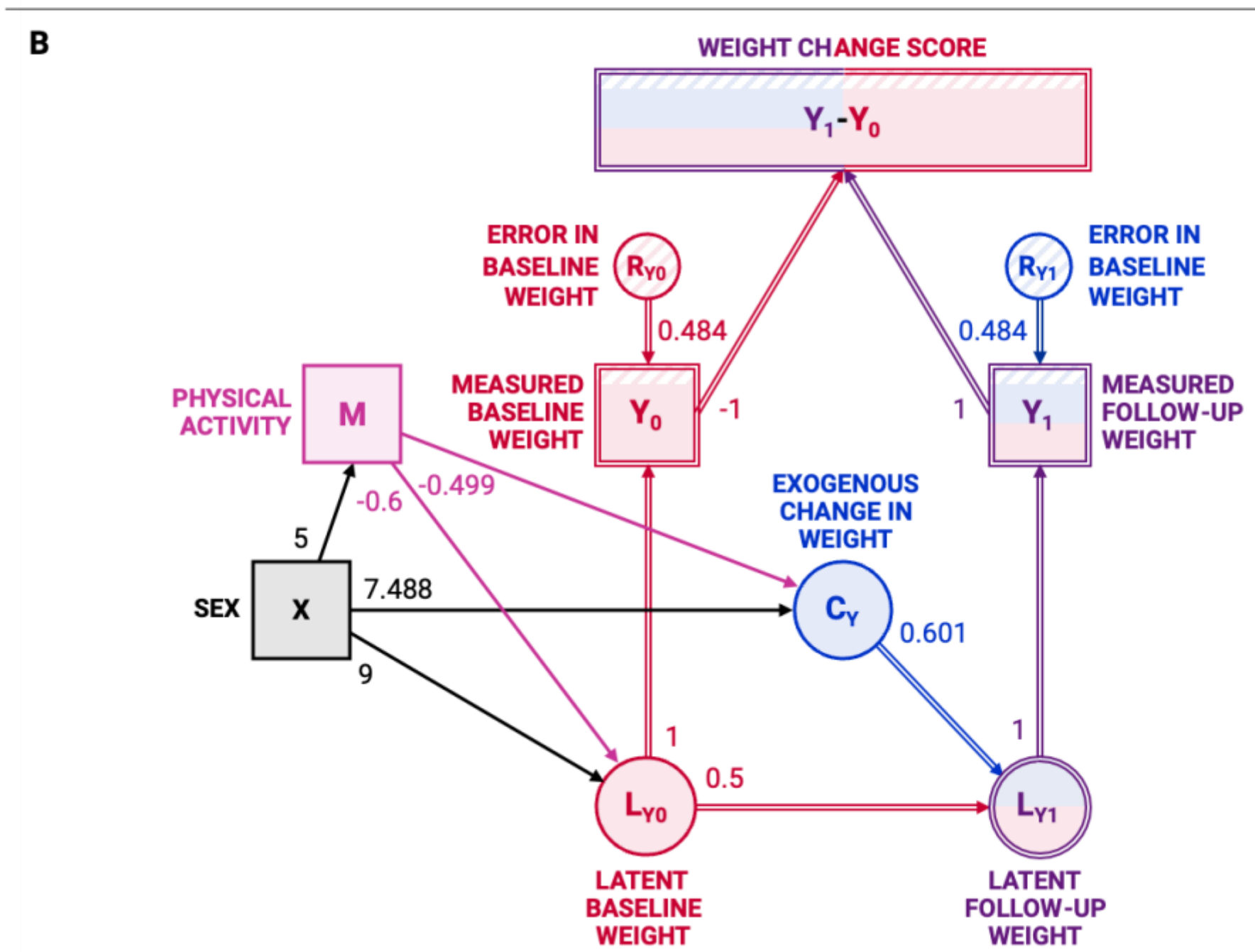

*$X_0$ represents the exposure, sex; $L_{Y0}$ and $Y_0$ are respectively the 'true' and 'measured' versions of the outcome, weight, at baseline, and $R_{Y0}$ represents all sources of transient random variation in baseline weight, such as measurement error. $C_Y$ represents all sources of exogenous change in weight between baseline and follow-up. $L_{Y1}$ and $Y_1$ are respectively the 'true' and 'measured' versions of the outcome, weight, at follow-up, and $R_{Y1}$ represents all sources of transient random variation in follow-up weight; $M_0$ represents a time-varying confounder, physical activity; $Y_1 - Y_0$ represents the change score derived from the 'measured' versions of the outcome at baseline and follow-up; As in **Figure 2**, measurable variables are depicted as squares, latent variables are depicted as circles, and the derived change score variable is depicted as a rectangle spanning from the baseline measurement period to the follow-up measurement period to represent the fact that it crystallises jointly across both timepoints (since it contains information from both). Variables depicted with double outlines (e.g., $Y_1$) are fully determined by their ancestors and have no additional sources of variation.[34]*

## Supplementary figure S2

Scatterplot of the simulated data for the scenario with time-varying confounding in the style of Lord's original diagram, showing the 'measured' follow-up weight ($Y_1$) plotted against the 'measured' baseline weight ($Y_0$), for boys and girls separately.

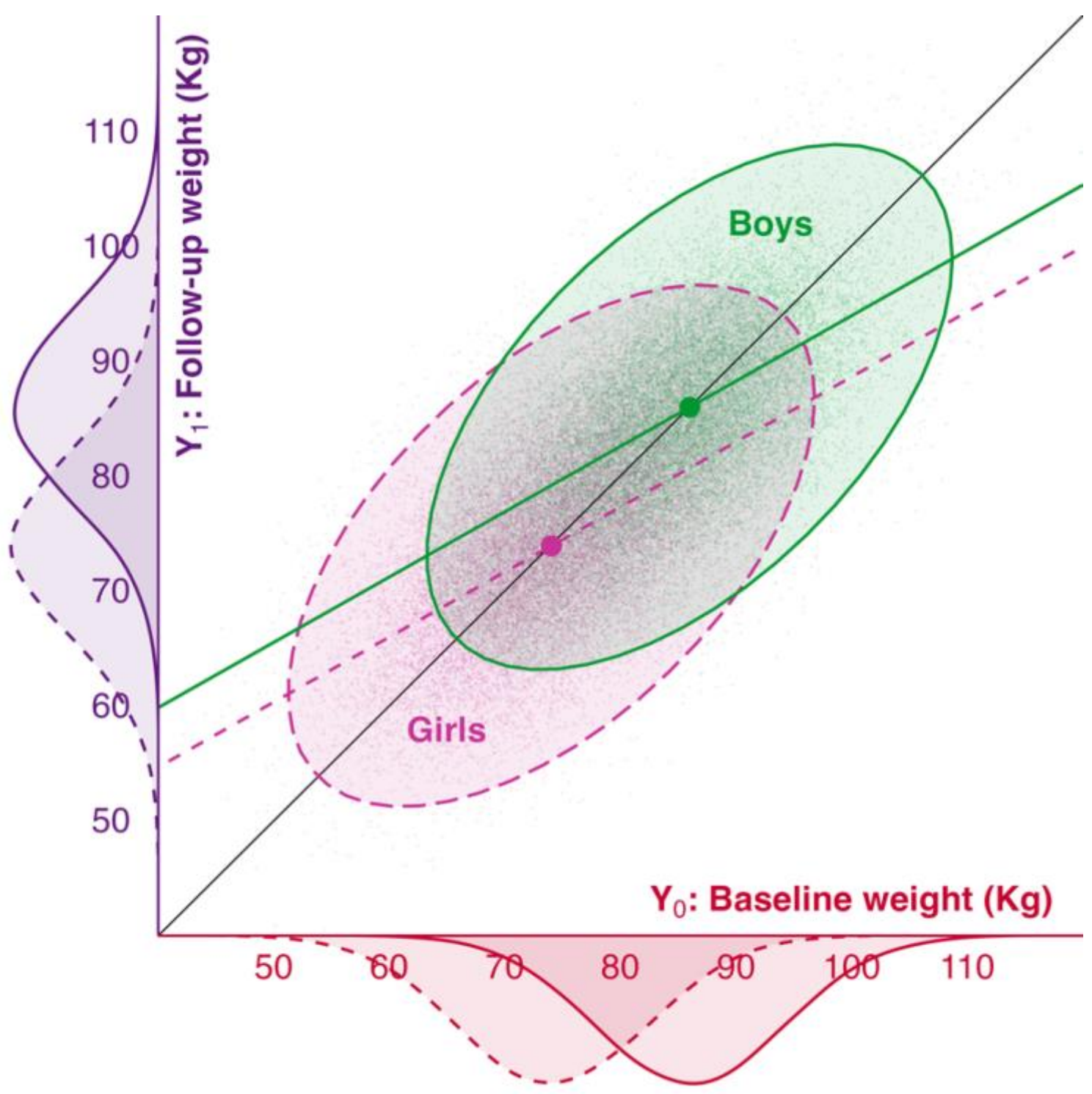

*The data for this plot comes from the second simulation scenario, where physical activity serves as a time-varying confounder for the controlled direct effect of sex on follow-up weight not mediated through baseline weight. See the **Figure 4** legend for a description of the features of this plot.*

## Supplementary references

35. Shi B, Choirat C, Coull BA, VanderWeele TJ, Valeri L. CMAverse: A Suite of Functions for Reproducible Causal Mediation Analyses. *Epidemiology*. 2021;32(5):e20.
36. Wickham H. ggplot2: Elegant Graphics for Data Analysis. New York, New York: Springer-Verlag; 2016.
37. Fox J, Weisberg S. Multivariate linear models in R. *An R Companion to Applied Regression.* Los Angeles, California: Thousand Oaks; 2011.